\documentclass[a4paper,11pt]{article}
\usepackage{jheppub} 
\usepackage{lineno}

\title{\boldmath Spectator partons' contribution in the rapidity width of $\Lambda$ in p+p collisions at various SPS and LHC energies}

 \author[1, 2]{Nur Hussain,}
\author[1]{Banajit Barman}
 \author[1]{ and Buddhadeb Bhattacharjee }

  \affiliation[1]{Nuclear and Radiation Physics Research Laboratory, Department of Physics, Gauhati University, Guwahati-781014, Assam, India}
   \affiliation[2]{Department of Physics, Madhab Choudhury College, Barpeta-781301, Assam, India}

\emailAdd{nur.hussain@cern.ch, banajit.barman@cern.ch, buddhadeb.bhattacharjee@cern.ch}

\abstract{Pseudorapidity distributions of all primary charged particles produced in p+p collisions at various Relativistic Heavy Ion Collider (RHIC) and Large Hadron Collider (LHC) energies using UrQMD-3.4 and PYTHIA8.2-generated events are estimated and compared with the existing results of UA5 and ALICE collaborations. With both the sets of generated data, the variation of rapidity widths with the rest masses of different mesons and baryons of p+p collisions at various Super Proton Synchrotron (SPS) and LHC energies are presented. An increase in the width of the rapidity distribution of $\Lambda$, similar to heavy-ion data, could be seen in p+p collisions as well from SPS to the LHC energies (up to $\sqrt{s}=13$ TeV) when the entire rapidity space is considered, indicating its universal feature both in p+p and A+A collisions. The genesis of the increased width of rapidity distribution of $\Lambda$ in p+p collision is found to be in the production of light flavored quarks and diquarks from the spectator partons of the colliding hadrons resulting $\Lambda$/$\bar{\Lambda}>1$ at the extreme rapidities of the studied p+p collisions. At the LHC energies for central p+p collision, when there is no spectator partons, the rapidity width of $\Lambda$ is found to take the same Gaussian shape as that of $\bar{\Lambda}$ and the widths of both the distributions become the same confirming no non-trivial contribution to the rapidity width of  $\Lambda$ as seen in heavy-ion and non-central p+p collisions. }


\begin{document}
\maketitle
\flushbottom

\section{Introduction}

Estimation of the widths of the rapidity distributions of identified charged particles of high energy nuclear collisions is of quite significance as they believe to carry a number of information about the dynamics of such collisions \cite{Klay, Bedanga, Blume, Deysir, nursir}. A systematic study on the variation of the widths of the rapidity distributions of various identified particles with their rest masses at different beam rapidities from AGS to SPS energies \cite{Blume}, reveals a jump in the rapidity width of $\Lambda$ with both UrQMD-generated and available experimental data \cite{Deysir}, resulting a separate mass scaling for mesons and baryons \cite{Deysir}. Such a jump in the rapidity width of $\Lambda$ was attributed to the net baryon density distribution effect \cite{Deysir}. Production of $\Lambda (uds)$ having two leading quarks, and not $\bar{\Lambda}$ ($\bar{u}\bar{d}\bar{s}$) having all produced quarks, is influenced by the net baryon density distribution, as a considerable fraction of $\Lambda$ at these collision energies are due to associated production \cite{Deysir2, AdlerAssociate, BassAssociate}. Subsequently, when this study was extended with both UrQMD generated and experimental data at SPS \cite{nursir, Blume, Afanasiev, Alt2, Alt}, and with UrQMD-generated data at RHIC and LHC energies, where  the collisions are much more transparent, but $B-\bar{B}$ is still greater than zero, this jump in the width of the rapidity distribution of $\Lambda$ is still found to exists and is considered to be a universal feature of heavy-ion collision data from the AGS and SPS to RHIC and LHC energies. Thus, the width of the rapidity distribution of $\Lambda$ produced in heavy-ion collisions is found to have a non-trivial non-kinematic contribution due to its associate production. In A+A collisions, up to the highest available LHC energy of RUN 2, with UrQMD-generated data, a situation could never been reached with $B-\bar{B} = 0$ and thus it could not be ascertained if the rapidity width of $\Lambda$ is free from net baryon density distribution effect at such energy.

It has been reported in Ref.~\cite{ALICEnetbaryon} by ALICE collaboration that $\bar{\Lambda}$/$\Lambda$ ratio becomes close to unity in p+p collisions at 7 TeV at mid-rapidity. It may, therefore, be expected that for such situation $B-\bar{B} $ would become zero and thus the $\Lambda$-production could be free from net baryon density effect. Here it is worth mentioning that at the LHC energies, the colliding protons are not considered merely as the collection of three valence quarks only. Such accelerated protons are rather believed to have a rich structure of many partons (gluons and sea quarks); when they collide, produce an assembly of large number of visible quarks and diquarks \cite{HM, lunddiquarks, TorbFragmantation}. These quarks and diquarks then evolve through various phases of space-time, suffer (multiple) scattering among themselves, and eventually resulted into visible hadrons through recombination and/or fragmentation mechanism resulting a situation similar to h-A or peripheral A-A collision of similar energies \cite{nature,ppHigh1, ppHigh2, ppHigh3, ppHigh4, ppHigh5, ppHigh6}. Thus, proton-proton collisions at the LHC energies can be considered as the collision of two composite structures resulting events of different multiplicity classes \cite{HM}.  However, one significant difference is that in A+A collisions, the spectator regions are composed of hadronic matter, whereas in p+p collisions the spectators are made up of partonic matter only. As it has been seen in Refs.~\cite{Deysir, nursir} for heavy-ion collisions that the spectator hadrons play a significant role in particle production, particularly in $\Lambda$ production, it would be interesting to see how these spectator partons of p+p collisions play its role in particle production. 

 It has been claimed in a number of reports  \cite{Bass1, Bleicher1, Bleicher2} that UrQMD is quite successful in describing heavy-ion collisions data over a wide range of energies. In Ref.~\cite{Mitrovski} it has also been shown that UrQMD  is equally successful in describing the experimental results of p+p, Au+Au, and Pb+Pb collisions from 17.3 GeV at the SPS to 1.8 TeV at Fermilab. On the other hand, PYTHIA has been found to be quite successful in explaining the p+p experimental results particularly at higher energies like that of LHC \cite{ALICECompare,ALICECompare2}.  Since UrQMD \cite{Bass} includes PYTHIA  to describe pp collisions, it is, therefore, expected that UrQMD should be successfully applied for small system like pp collisions at LHC energies as well. In this work an attempt has therefore been made, with UrQMD-3.4 and PYTHIA8.2 (4C and Monash tuned) generated p+p events at various colliding energies from SPS to the LHC (up to $\sqrt{s}=13$  TeV), to estimate the widths of the rapidity distributions of various produced particles at different beam energies to examine  the role of spectator partons on rapidity width of $\Lambda$ and hence on $\Lambda$ ($\bar{\Lambda}$) production mechanism.

\section{MC event generators}
For the present study, we generated 30 to 80 millions p+p inelastic events at each studied energies using UrQMD-3.4, PYTHIA8.2 4C and Monash tunings. UrQMD is a many-body microscopic Monte Carlo event generator based on the covariant propagation of color strings, constituent quarks and diquarks with mesonic and baryonic degrees of freedom  \cite{Bass, Bleicher, Petersen, Andersson}. It includes PYTHIA to incorporate the hard perturbative QCD effects \cite{Bleicher, Bass}. The current version of the UrQMD model also includes the excitation and fragmentation of color strings, formation and decays of hadronic resonances, and rescattering of particles. On the other hand, PYTHIA, a standalone event generator, consists of a coherent set of physics models, which describes the evolution of high-energy collisions from a few-body hard-scattering processes to a complex multiparticle final state \cite{Torbjorn1,Torbjorn2,Torbjorn3, Sjostrand}. 
In PYTHIA, all main aspects of events such as, hard processes selection, initial and final state radiation, beam remnants, fragmentation, decays, etc. are simulated. The model generated events should, in principle, be comparable with the experimental observables of ultra-relativistic nuclear collision and should be able to predict physics information from such comparison or to study physics at future experiment  \cite{Torbjorn2, Torbjorn3}. The available high precision experimental data allow detailed model comparisons and motivate the effort on model development and tuning of the existing models towards more precise predictions. One of the dedicated tunings of PYTHIA8 event generator is PYTHIA8 4C, which is used to predict the results of Run 1 LHC data and the modified parameters are reported in Ref.~\cite{Pythia4C}. As reported in Ref.~\cite{PythiaMonash}, the PYTHIA8 version is tuned with Monash, which has been constructed to give a reasonable description of soft inclusive (minimum bias) and underlying-event type observables of LHC Run 2 data. Further, Monash tune includes multi-parton interactions and color reconnection profile for hadron-hadron collision and all other tuning parameters are reported in Ref.~\cite{PythiaMonash}. 

\section{Results}
\begin{figure}[htbp]
		\centering
		\includegraphics[width=12cm, height=8 cm]{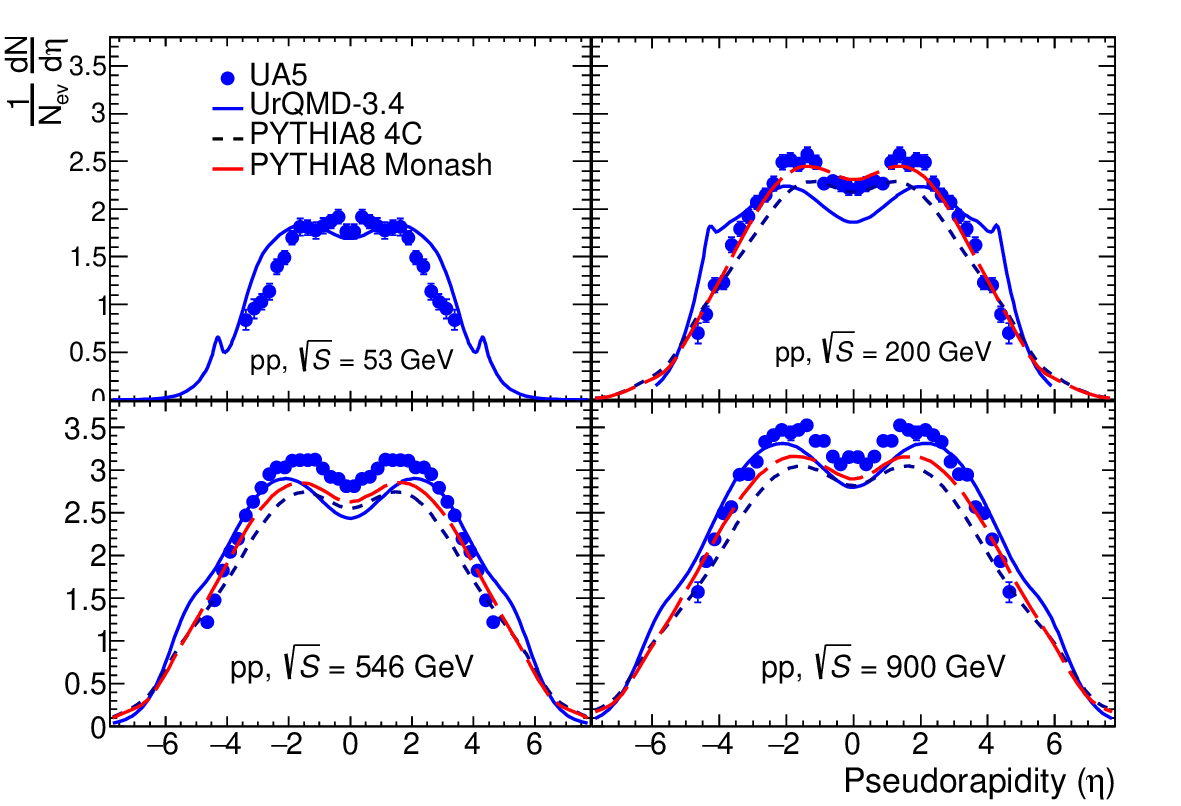}
		\caption{Comparison of pseudorapidity distributions of all primary charged particles with UrQMD-3.4 and PYTHIA8.2 (4C and Monash) generated events in p+p collisions at $\sqrt{\it{s}} = 53, 546, 200$, and $900$ GeV with UA5 collaboration \cite{AlnerUA5}.}
		\label{Fig:etaUA5}	
\end{figure}

\begin{figure}[htbp]
		\centering
		\includegraphics[width=12cm, height=8 cm]{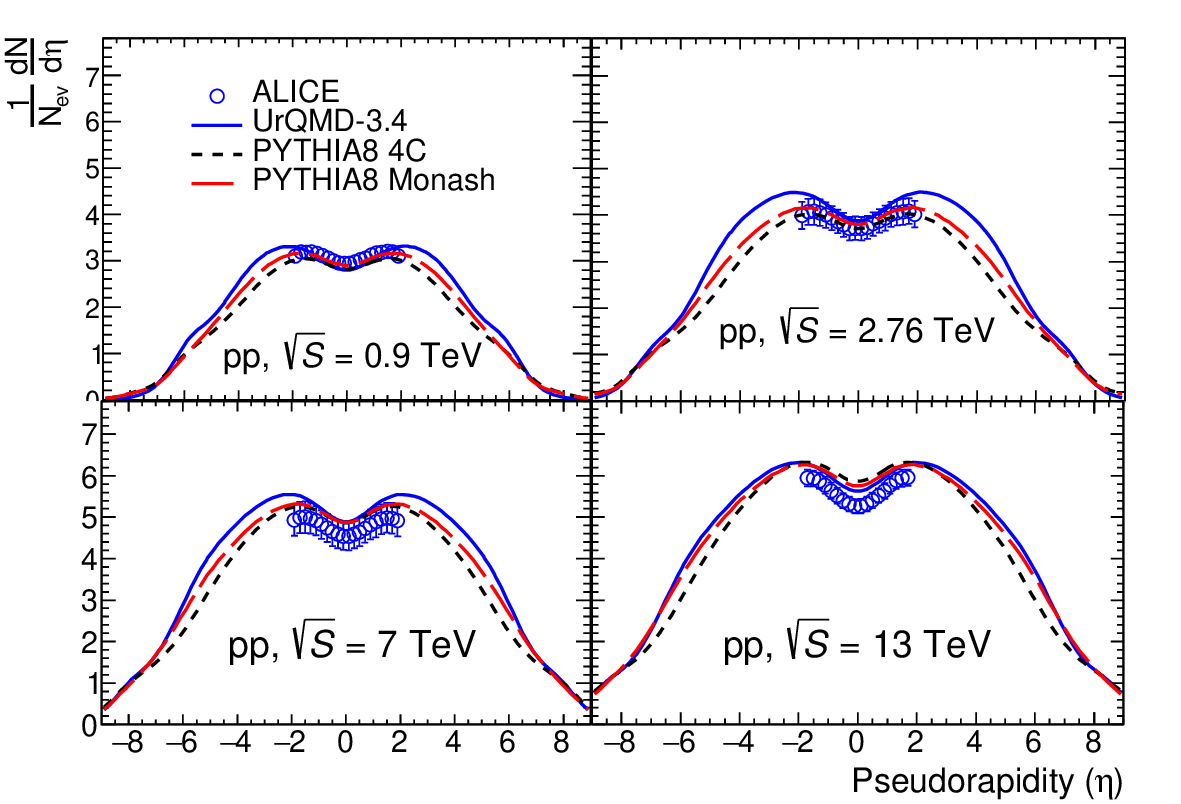}
		\caption{Comparison of pseudorapidity distributions of all primary charged particles with UrQMD-3.4 and PYTHIA8.2 (4C and Monash) generated events in p+p collisions at $\sqrt{\it{s}} = 0.9, 2.76, 7,$ and $13$ TeV with ALICE collaboration \cite{ALICECompare,ALICECompare2}.}
		\label{Fig:etaALICE}	
\end{figure}

 In our earlier work \cite{nursir}, it has been shown that UrQMD-3.4-generated (pseudo)rapidity distributions of produced particles show a good agreement with the experimental results at all the SPS, RHIC, and LHC energies for Au+Au and Pb+Pb systems. With the present sets of generated data, pseudorapidity distributions of all primary charged particles are compared with the various existing experimental results of p+p collisions of UA5 and ALICE collaborations. Pseudorapidity distributions of all primary charged particles using UrQMD-3.4 and PYTHIA8.2-generated events of p+p collisions at $\sqrt{\it{s}} = 53$ (only UrQMD), $200$, $546$, and $900$ GeV are compared with the results of UA5 collaboration \cite{AlnerUA5} and is shown in Fig.~\ref{Fig:etaUA5}. Though the MC data reproduces well the shape of the experimental pseudorapidity distributions of UA5 collaboration, around mid-rapidity, the models somewhat under-predict the experimental values at all the studied energies of UA5 collaboration. However, in Fig.~\ref{Fig:etaALICE}, all sets of generated data (UrQMD-3.4, PYTHIA8.2 4C, and PYTHIA8.2 Monash) at $\sqrt{\it{s}} = 900$ GeV show a better agreement with the experimental results of ALICE collaboration \cite{ALICECompare,ALICECompare2}. Further details of comparison of model generated and experimental data could be found in Refs.~\cite{ALICECompare,ALICECompare2, kaidalovSingleDifrative, UA5inconsistenceData}. Considering the fact that all the three sets of model-predicted pseudorapidity distributions are in good agreement with the experimental results of appropriate energies, further analysis is carried out with all the three, i.e., UrQMD-3.4, PYTHIA8.2 4C, and PYTHIA8.2 Monash model-generated data. 
 
\begin{figure}[htbp]
		\centering
		\includegraphics[width=11cm, height=7 cm]{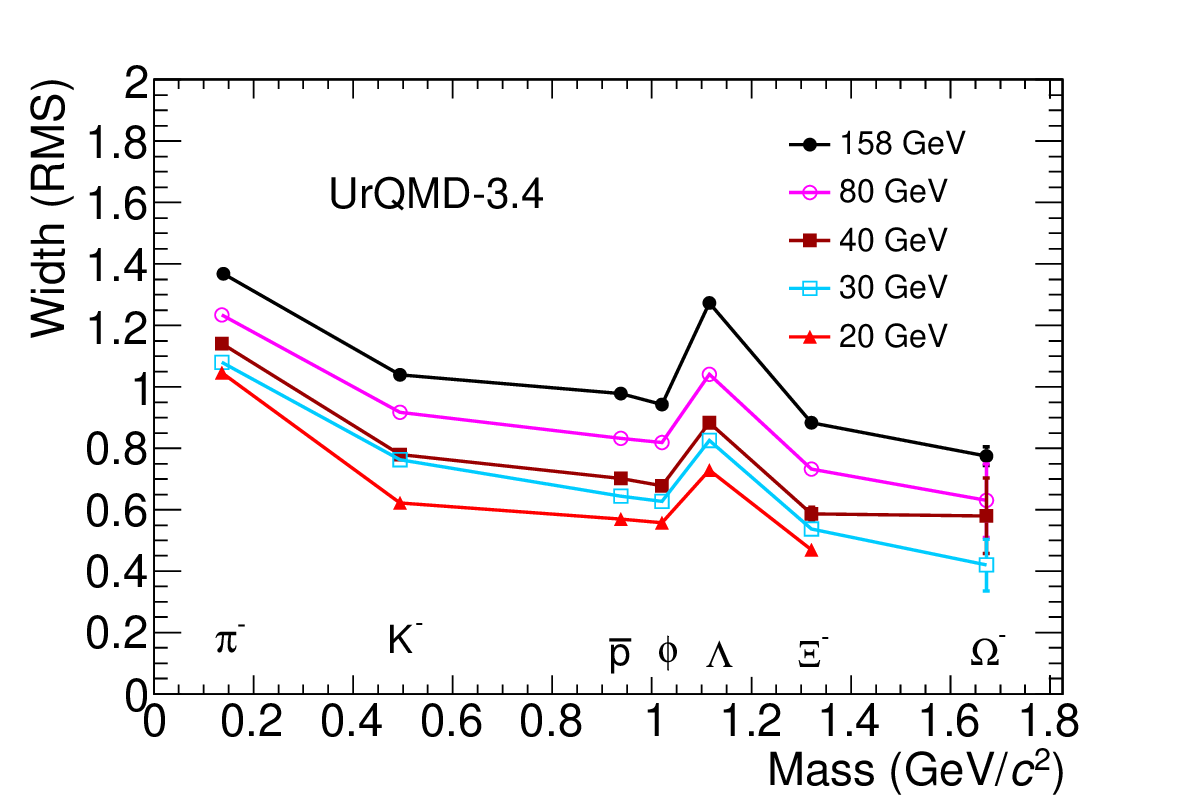}
		\caption{Widths of the rapidity distributions of various produced particles as a function of their rest masses in p+p collisions at all the SPS energies with UrQMD-3.4-generated events.}
		\label{Fig:widthSPS}	
\end{figure}

\begin{figure}[htbp]
		\centering
		\includegraphics[width=11cm, height=7 cm]{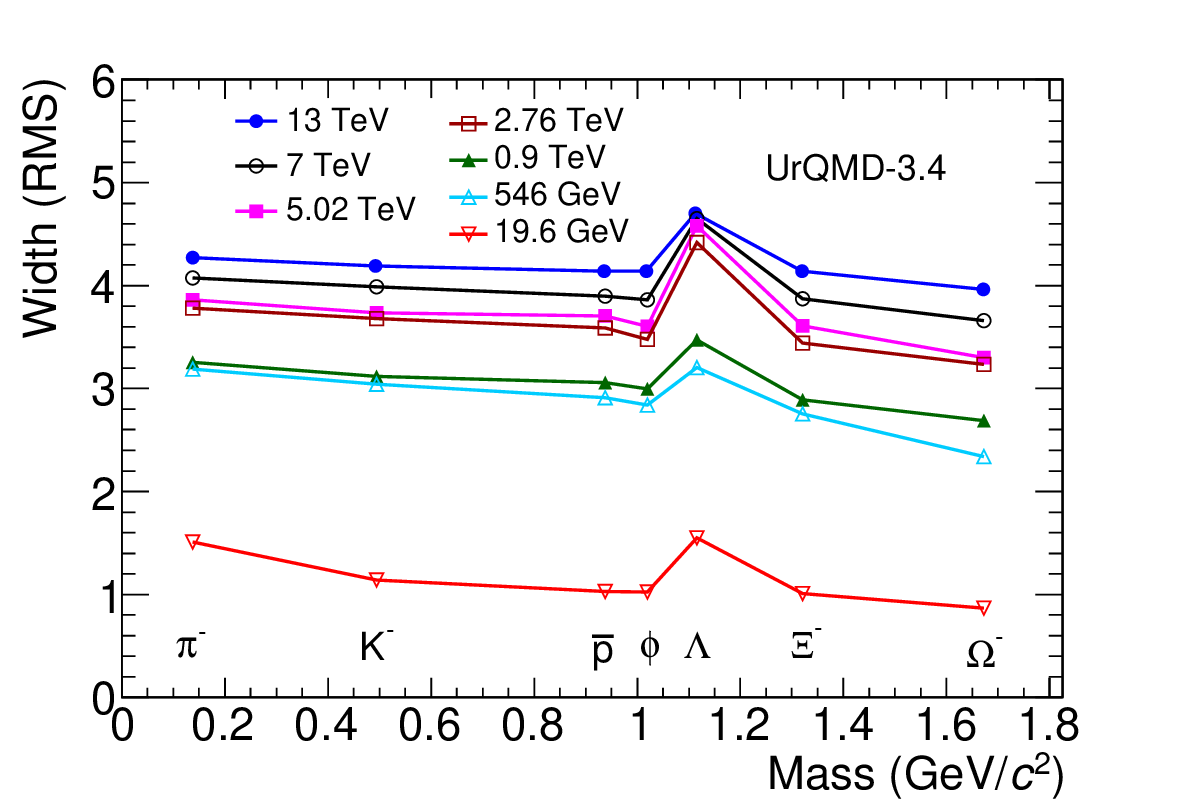}
		\caption{Widths of the rapidity distributions of various produced particles as a function of their rest masses in p+p collisions at lower RHIC and all the studied LHC energies with UrQMD-3.4-generated events for $|y| \leq 6.5$.}
		\label{Fig:widthLHC}	
\end{figure}
\begin{figure}[htbp]
		\centering
		\includegraphics[width=11cm, height=7cm]{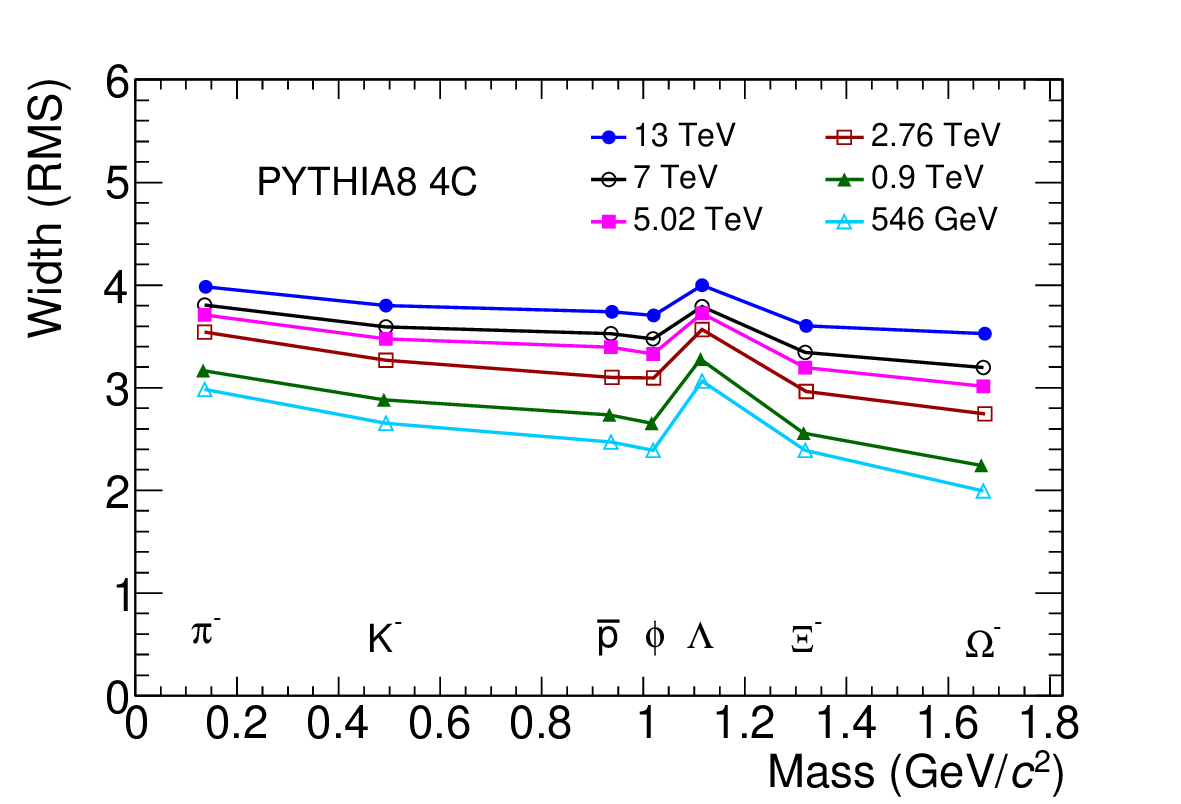}
		\caption{Widths of the rapidity distributions of various produced particles as a function of their rest masses in p+p collisions from $\sqrt{\it{s}} = 0.546 $ TeV to 13 TeV with PYTHIA8.2 4C-generated inelastic events for $|y| \leq 7$.}
		\label{Fig:widthPythia4c}	
\end{figure}
\begin{figure}[htbp]
		\centering
		\includegraphics[width=11cm, height=7cm]{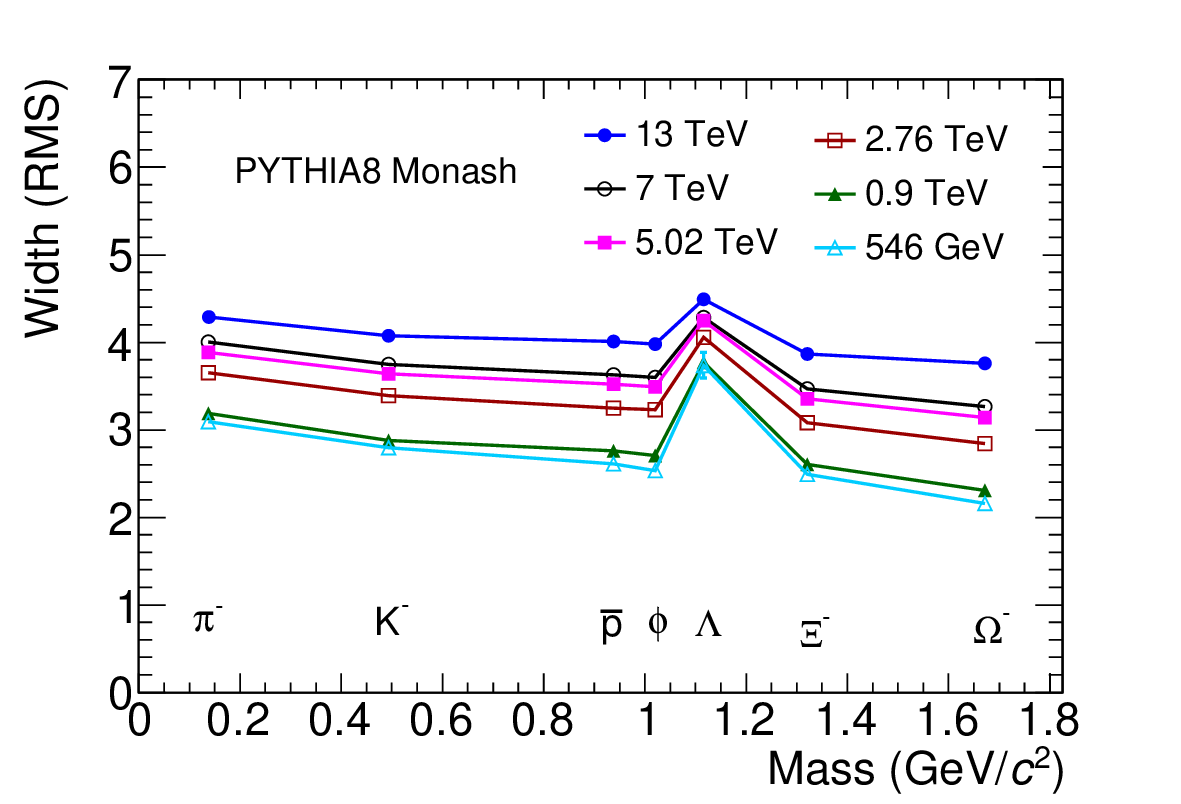}
		\caption{Widths of the rapidity distributions of various produced particles as a function of their rest masses in p+p collisions from $\sqrt{\it{s}} = 0.546 $ TeV to 13 TeV with PYTHIA8.2 Monash-generated inelastic events for $|y| \leq 7$.}
		\label{Fig:widthPythiaMonash}	
\end{figure}
UrQMD-3.4- and PYTHIA8.2 (4C and Monash)-generated rapidity distributions of $\pi^-,  K^-, p(\bar{p}), \phi $, $ \Lambda (\bar{\Lambda})$, $\Xi^-$, and $\Omega^-$ at all the studied energies for p+p collisions are parameterized by the following double Gaussian function \cite{Deysir,Alt}, 
\begin{equation} \label{eq:1}
\frac{dN}{dy} \propto ( \, e^{- \frac{(y-\bar{y})^2}{2\sigma^2}} + e^{- \frac{(y+\bar{y})^2}{2\sigma^2}}), \,\\
\end{equation}
where the symbols have their usual significance. Using this fitting function, widths of the rapidity distributions of generated data of all the studied hadrons are estimated and listed in Tables~\ref{tableSPSWidth} and \ref{tableSPSWidthPythia}. The variation of the rapidity widths of various mesons and baryons as a function of their rest masses at all the SPS, lower RHIC, and all the LHC energies for the rapidity space $|y| \leq7$ are shown in Figs.~\ref{Fig:widthSPS}, \ref{Fig:widthLHC}, \ref{Fig:widthPythia4c}, and \ref{Fig:widthPythiaMonash}. 

\begin{figure}[htbp]
		\centering
		\includegraphics[width=12 cm,height=6cm]{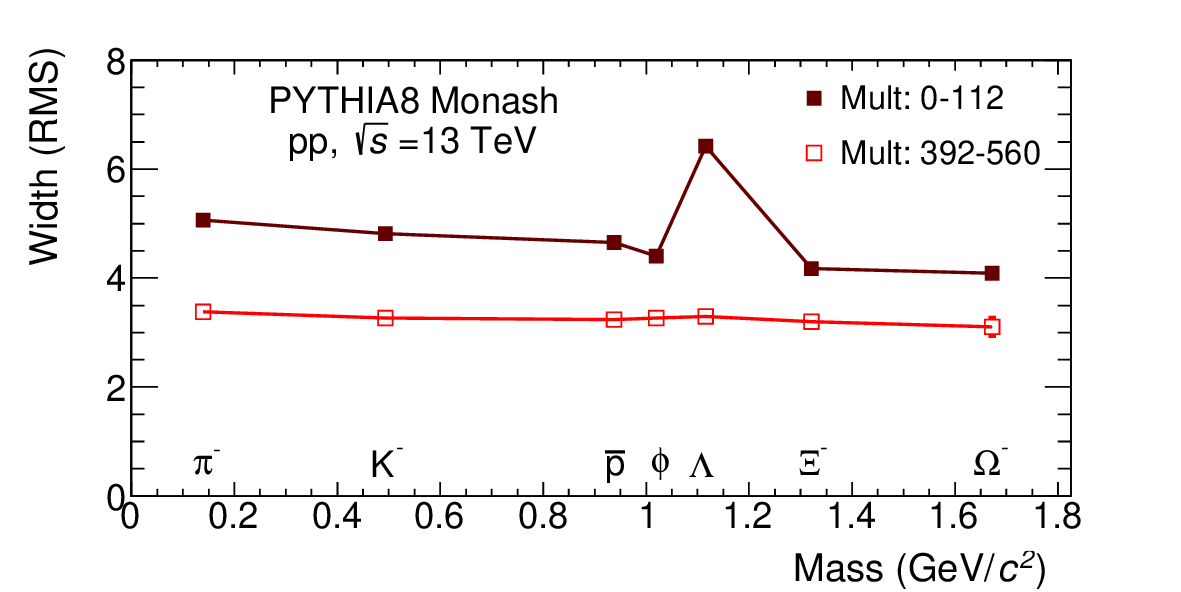}
		\caption{Widths of the rapidity distributions of various produced particles as a function of their rest masses in p+p collisions at $\sqrt{s} = 13$ TeV with PYTHIA8.2 Monash-generated inelastic events for low ($0-112$) and high ($392-560$) multiplicity classes.}
		\label{Fig:WidthMultiplicity}	
\end{figure}

\begin{center}
\setlength{\tabcolsep}{0.5em}
\begin{table*}[htbp]
\caption {Widths of the rapidity distributions of all the studied hadrons with UrQMD-3.4-generated p+p events at all the SPS, lower RHIC, and all the LHC energies. At LHC energies, rapidity widths of $\Lambda$s are estimated using  double Gaussian function excluding the extreme two peaks, whereas rapidity widths estimated using triple Gaussian function including all the three peaks are shown in the parentheses.}
 \label{tableSPSWidth}
\centering
\resizebox{\textwidth}{!}{%
\begin{tabular}{ c c c c c c c c}
\hline
\hline
 Energy                       &$\pi^-$                               &   $K^-$                       & $\bar{p}$                          & $\phi$                             &  $\Lambda$                      & $\Xi^-$                      & $\Omega^-$         \\ 
  $\sqrt{\it{s}}$ ($E_{lab}$)&& &  &  &  &   &  \\
\hline
 ($20$ GeV)                 &$1.045\pm0.001$	          &$0.622\pm0.002$      &$0.570\pm0.003$          &$0.558\pm0.011$            &$0.728\pm0.001$         & $0.468\pm0.017 $	& -    \\
 ($30 $ GeV)                & $1.078\pm0.001$         &$0.762\pm0.002$	     &$0.644\pm0.003$      &$0.627\pm0.007$	             &$0.826\pm0.001$	   &$0.536\pm0.009$	         &$0.419\pm0.084$ \\                                           
 ($40$ GeV)	           & $1.115\pm0.002$	  &$0.779\pm0.006$	      & $0.701\pm0.008$      &$0.678\pm0.017$               &$0.883\pm0.002$ 	   &$0.586\pm0.022$        &$0.590\pm0.240$ \\
 ($80$ GeV)		   &$1.230\pm0.001$ 	  &$0.916\pm0.007$	      &$0.832\pm0.005$       &$0.818\pm0.014$              &$1.040\pm0.002$	   &$0.731\pm0.014$        &$0.630\pm0.120$  \\
 ($158$ GeV)               & $1.369\pm0.001$          &$1.039\pm0.002$	      & $0.979\pm0.003$     &$0.942\pm0.008$	             &$1.273\pm0.001$ 	   &$0.883\pm0.010$	        &$ 0.774\pm0.029$ \\
$19.6$ GeV	            & $1.414\pm0.001$	   &$1.137\pm0.002$      &$1.031\pm0.002$ 	  &$0.993\pm0.005$	             &$1.352\pm0.001$	   &$0.963\pm0.006$ 	& $0.833\pm0.017$ \\
$546$ GeV	            & $3.174\pm0.002$          &$3.043\pm0.007$       & $2.913\pm0.007$	  &$2.842\pm0.013$	             &$3.204\pm0.024$	   &$ 2.754\pm0.035$ 	 &$2.337\pm0.014$ \\
 $900$ GeV	            & $3.254\pm0.002$          &$3.122\pm0.006$	       &$3.056\pm0.002$ 	  &$2.989\pm0.014$	             &$3.487\pm0.036$	   &$2.909\pm0.042$ 	 &$2.461\pm0.093$ \\   
 				    &					   &				      & 	                          &					     &($4.55\pm0.019$)          &					 &				\\
 $2.76$ TeV	            & $3.766\pm0.001$           &$3.681\pm0.001$      &$3.587\pm0.002$ 	  &$3.479\pm0.002$              &$4.417\pm0.014$	   &$3.444\pm0.007$	         &$3.237\pm0.030$ \\ 
 			            &				             &				      & 	                           &					     &($5.091\pm0.003$)         &					 &				\\ 
  $5.02$ TeV	           & $3.861\pm0.002$            &$3.735\pm0.002$      &$3.704\pm0.002$ 	  &$3.604\pm0.004$                &$4.583\pm0.028$	   &$3.611\pm0.015$         &$3.301\pm0.065$  \\ 
   				   &					    &				       & 	                            &					     &($5.152\pm0.008$)        &					 &				\\ 
  $7$ TeV	                    & $4.060\pm0.001$            &$3.987\pm0.001$	 &$3.897\pm0.001$     &$3.860\pm0.003$	              &$4.650\pm0.013$ 	   &$3.871\pm0.010$	         &$ 3.662\pm0.043$  \\
   			           &					     &				         & 	                             &					     &($5.158\pm0.003$)       &					 &				\\ 
  $13$ TeV	           &$4.267\pm0.001$             &$4.178\pm0.001$	  &$4.138\pm0.001$      &$4.129\pm0.003$              &$4.690\pm0.009$  	   &$4.140\pm0.012$        &$3.953\pm0.051$ \\ 
     					   &				    &				          & 	                              &					     &($5.180\pm0.003$)      &					 &				\\ 
\hline
\hline
\end{tabular}}
\end{table*}
\end{center}
\begin{center}
\setlength{\tabcolsep}{0.5em}
\begin{table*}[htbp]
\caption {Widths of the rapidity distributions of all the studied hadrons with PYTHIA8.2 Monash-generated inelastic p+p events from 546 GeV to the top LHC energies. At LHC energies, rapidity widths of $\Lambda$s are estimated using  double Gaussian function excluding the extreme two peaks, whereas rapidity widths estimated using triple Gaussian function including all the three peaks are shown in the parentheses.}
 \label{tableSPSWidthPythia}
\centering
\resizebox{\textwidth}{!}{%
\begin{tabular}{ c c c c c c c c}
\hline
\hline
 Energy                                 &$\pi^-$                                 &   $K^-$                  &$\bar{p}$              & $\phi$                             &  $\Lambda$                      & $\Xi^-$                      & $\Omega^-$          \\ 
  $\sqrt{\it{s}}$ && &  &  &  &   &  \\
\hline
 $546$ GeV	                    & $3.078\pm0.001$                    &$2.796\pm0.002$   &$2.609\pm0.001$	  &$2.533\pm0.003$	             &$3.738\pm0.144$	   &$ 2.490\pm0.007$ 	 &$2.156\pm0.020$   \\
 $900$ GeV	                    & $3.174\pm0.001$                   &$2.876\pm0.001$	   &$2.761\pm0.001$	  &$2.705\pm0.002$	             &$3.770\pm0.044$	   &$2.606\pm0.005$ 	 &$2.307\pm0.017$   \\   
 					   &						&				&    &					     &($4.352\pm0.010$)          &					 &				\\
 $2.76$ TeV	                    & $3.647\pm0.001$                   &$3.390\pm0.001$  &$3.248\pm0.001$	  &$3.227\pm0.003$              &$4.056\pm0.023$	   &$3.080\pm0.006$	         &$2.845\pm0.021$   \\ 
 					   &						&				&	&					     &($4.655\pm0.007$)         &					 &				\\ 
  $5.02$ TeV	                    & $3.880\pm0.001$                   &$3.639\pm0.001$ &$3.520\pm0.001$	  &$3.493\pm0.004$                &$4.248\pm0.015$	   &$3.352\pm0.005$         &$3.143\pm0.020$   \\ 
   					   &						&				&	&					     &($4.826\pm0.006$)        &					 &				\\ 
  $7$ TeV	                            & $3.985\pm0.001$                 &$3.748\pm0.001$	  &$3.631\pm0.001$ 	  &$3.599\pm0.002$	              &$4.283\pm0.011$ 	   &$3.468\pm0.005$	         &$ 3.267\pm0.017$   \\ 
   					   &						&				&	&					     &($4.925\pm0.005$)       &					 &				\\ 
  $13$ TeV	                    &$4.291\pm0.002$                  &$4.075\pm0.002$	  &$4.007\pm0.002$ 	  &$3.981\pm0.005$              &$4.489\pm0.011$  	   &$3.864\pm0.009$        &$3.762\pm0.037$   \\ 
     					   &						&				&	&					     &($5.085\pm0.006$)      &					 &				\\ 
\hline
\hline
\end{tabular}}
\end{table*}
\end{center}
\begin{figure}[htbp]
		\centering
		\includegraphics[width=12cm, height=8cm]{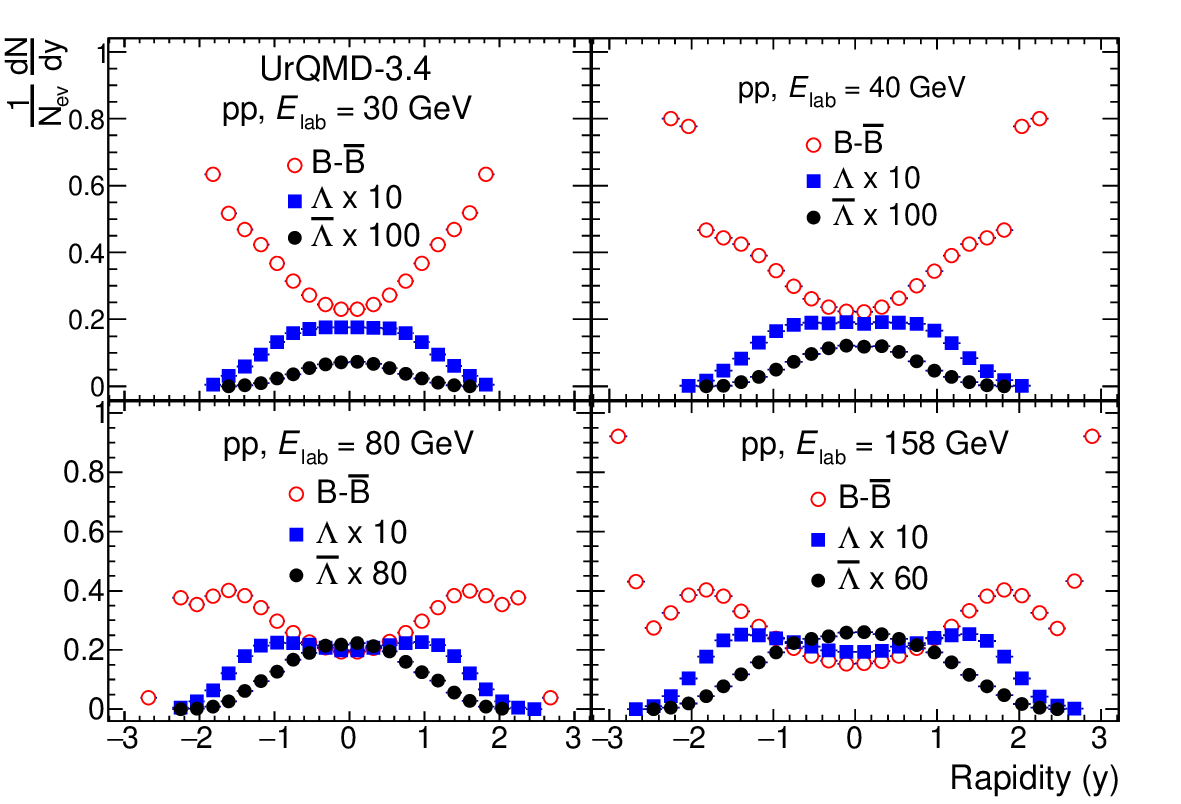}
		\caption{Distributions of $B-\bar{B}$, $\Lambda$, and $\bar{\Lambda}$ over the entire rapidity space with UrQMD-3.4-generated events in p+p collisions at various SPS energies.}
		\label{Fig:netbaryonSPS}	
\end{figure}
\begin{figure}[htbp]
		\centering
		\includegraphics[width=12cm, height=8 cm]{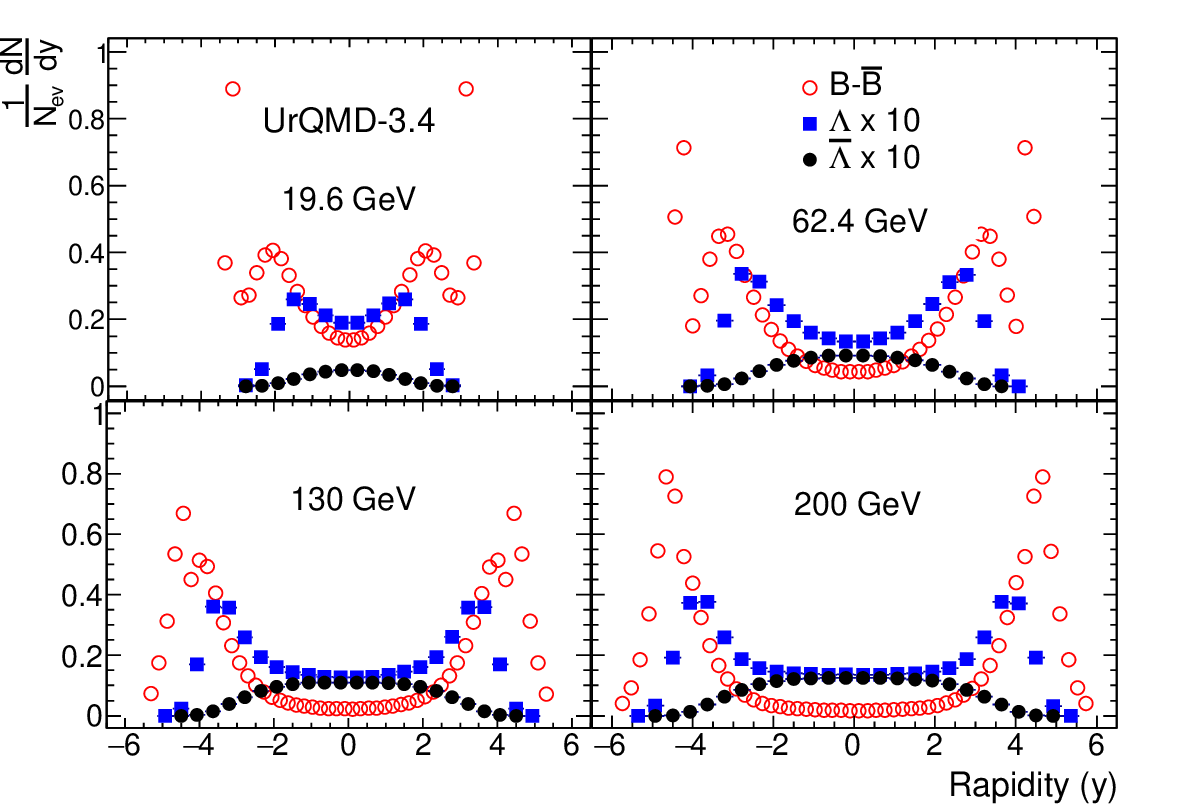}
		\caption{Distributions of $B-\bar{B}$, $\Lambda$, and $\bar{\Lambda}$ over the entire rapidity space with UrQMD-3.4-generated events in p+p collisions at various RHIC energies.}
		\label{Fig:netbaryonRHIC}	
\end{figure}
It could be readily seen from Figs.~\ref{Fig:widthSPS}, \ref{Fig:widthLHC},  \ref{Fig:widthPythia4c}, and \ref{Fig:widthPythiaMonash} that, as reported in Refs.~\cite{Deysir,nursir} for heavy-ion collisions from AGS to LHC energies, a similar jump in the rapidity width of $\Lambda$ exists also in the case of minimum biased p+p collisions from low SPS to the LHC energies (up to $\sqrt{s}=13$ TeV) for both UrQMD-3.4 and PYTHIA8.2 (4C and Monash tuned) models generated data. Such an increase in the width of the rapidity distribution of $\Lambda (uds)$ and not   $\bar{\Lambda}$ ($\bar{u}\bar{d}\bar{s}$) in A+A collision was attributed to the dependence of $\Lambda$  production on net baryon density distribution or otherwise on associated production of $\Lambda$  \cite{Deysir,nursir}. In p+p collisions, as shown in Figs.~\ref{Fig:netbaryonSPS} and \ref{Fig:netbaryonRHIC}, the net baryon density, as in the case of heavy-ion collision, is found to be maximum at extreme rapidities at SPS and RHIC energies and minimum but not zero at zero-rapidity. It is interesting to mention here that, unlike in heavy-ion collisions, the spectator regions of p+p collisions consist of partonic rather than hadronic matter and therefore the observed maximum net baryon density at extreme rapidity spaces in p+p collisions is not due to the leading hadrons but due to the hadrons produced out of partons of the spectators. 
\begin{figure}[htbp]
		\centering
		\includegraphics[width=14cm,height=6cm]{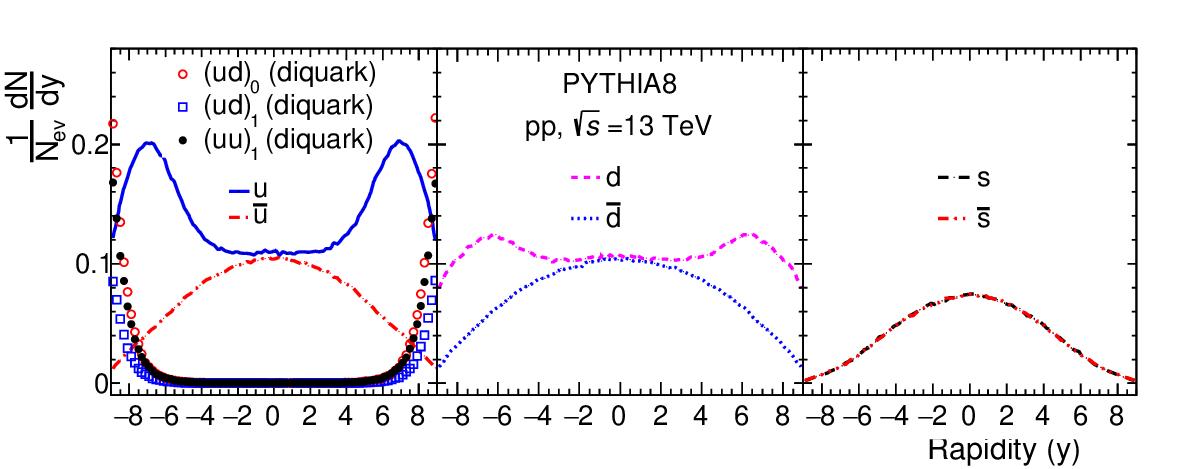}
		\caption{Rapidity distributions of $u, \bar{u}, d, \bar{d}, s$, $\bar{s}$ and various diquarks in p+p collisions at $\sqrt{\it{s}} = 13$ TeV with PYTHIA8.2 Monash-generated events. }
		\label{Fig:quarks}
\end{figure}
\begin{figure}[htbp]
		\centering
		\includegraphics[width=12cm, height=6cm]{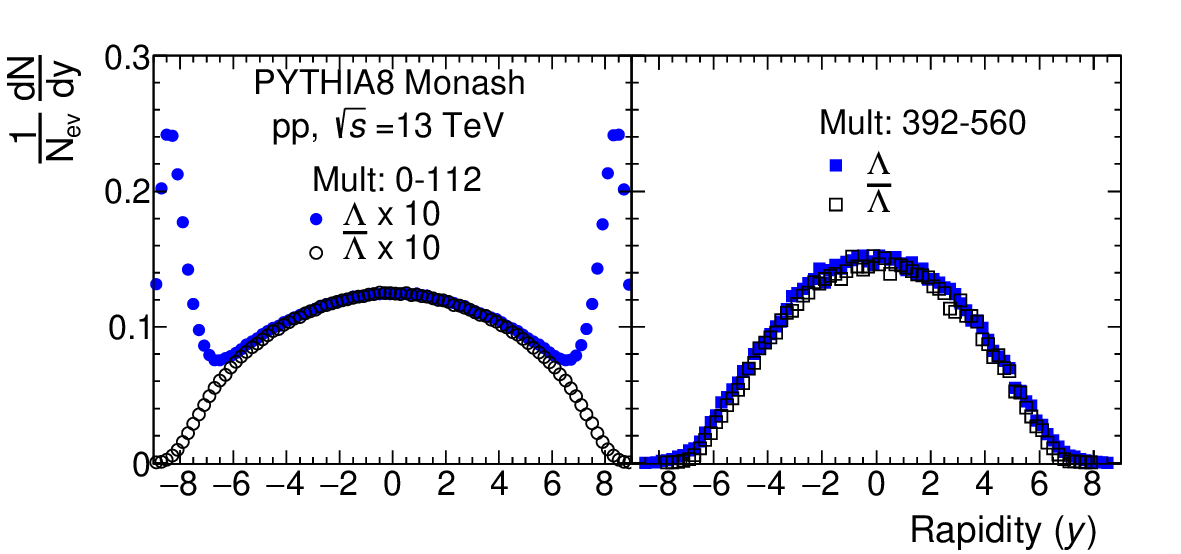}
		\caption{Rapidity distributions of $\Lambda$ and $\bar{\Lambda}$ in p+p collisions at $\sqrt{\it{s}} = 13$ TeV for lowest multiplicity (left panel) and highest multiplicity (right panel) classes with PYTHIA8.2 Monash-generated inelastic events.}
		\label{Fig:RapidityMult}	
\end{figure}
A high net baryon density at extreme rapidities of p+p collisions at studied energies indicates that the spectator partons (mostly, gluons and sea quarks) might have been resulted in the availability of more light-flavored quarks and diquarks than their counterparts, which eventually resulted in the production of more number of $\Lambda$ (baryons) than $\bar{\Lambda}$ (anti-baryons) resulting $\Lambda$/$\bar{\Lambda}>1$. To examine the validity of this argument, we have compared, in Fig.~\ref{Fig:quarks}, the rapidity distribution of $u$, $\bar{u}$, $d$, $\bar{d}$, $s$, $\bar{s}$ quarks and various diquarks over the entire rapidity space. It could be readily seen from these plots that both the light flavored $u$ and $d$ quarks  and various diquarks show a bump in their yields in the extreme rapidity regions. Moreover, it is interesting to note that the rapidity distribution of $s$ and $\bar{s}$ follow exactly the same pattern (extreme right panel of Fig.~\ref{Fig:quarks}) implying that light flavored quarks and diquarks are more abundant in the beam rapidity region than their counterparts. It can therefore be inferred without much dubiety that the rapidity distribution of $\Lambda$ and $\bar{\Lambda}$, whose constituent quarks are $uds$ and $\bar{u}\bar{d}\bar{s}$ respectively and which might have been produced from a diquark and $s$ quark, follow the same pattern as their constituent light-flavored diquarks (quark).  However, when there is no beam remnants (for the most central collision) the rapidity distributions of $\Lambda$ and $\bar{\Lambda}$ become same (Fig.~\ref{Fig:RapidityMult}), and thus, the jump in the rapidity width of $\Lambda$, as can be readily seen from Fig.~\ref{Fig:WidthMultiplicity}, almost disappears and the rapidity widths of $\Lambda$ and $\bar{\Lambda}$, within error, become equal. Thus, in pp collisions the jump in the width of rapidity distribution of  $\Lambda$, and not $\bar{\Lambda}$, occurs due to production of light flavoured quarks and diquarks at the extreme rapidities from the spectator partons of the colliding hadrons. At most central collision, when there is no spectator region, the widths of rapidity distributions of $\Lambda$  and $\bar{\Lambda}$  become the same. 

\section{Summary}
In this work UrQMD-3.4 and PYTHIA8.2 (4C and Monash) generated pseudorapidity distributions of all primary charged particles of p+p collisions are compared with the existing experimental results of UA5 and ALICE collaborations which show a good agreement between model generated and experimental results for the studied energies. An increase in the rapidity width of $\Lambda$, as observed in heavy-ion data \cite{nursir}, could be seen with the generated data of p+p collisions as well at all the SPS energies. When the study is extended to RHIC and LHC energies, a similar increase in the rapidity width of $\Lambda$ could be observed for the consideration of full rapidity space ($|y|\leq7$) indicating that the increase in the rapidity width of $\Lambda$ is a general characteristic of both p+p and A+A collisions from SPS to RHIC and LHC energies. While in A+A collisions, the observed increase in the width of the rapidity distribution of $\Lambda (uds)$ is attributed to their associated production from spectator hadrons, in p+p collisions it could be due to favourable production of light flavoured quarks and diquarks from the spectator regions of p+p collisions. Such favorable production of light flavored diquarks (and quarks) over their counterparts might resulted $\Lambda$ production different from mid rapidity regime.  That our consideration of production of $\Lambda$  at high rapidity spaces  due to production of light flavoured diquarks and quarks produced from the spectator partons is correct is vindicated by the facts that the rapidity distributions of $\Lambda$ and $\bar{\Lambda}$ for most central collision, when there is no spectator regions, follow the same Gaussian shape and the widths of the rapidity distribution of $\Lambda$ and $\bar{\Lambda}$ become the same. It could, therefore, be easily comprehended that there are some interesting physics in longitudinal direction as well, particularly in p+p collision, and need to be explored in future.

\section*{Acknowledgments}
The financial assistance of Department of Science and Technology (DST), Government of India through the project No. SR/MF/PS-02/2021-GU(E-37122) is thankfully acknowledged. The authors convey their thanks to the teams of UrQMD and PYTHIA for developing their codes and allowing these freely available on public domain.


\end{document}